\documentclass[pra,twocolumn,superscriptaddress,showpacs]{revtex4}
\usepackage{epsfig}
\usepackage{amsbsy}
\usepackage{amsmath}
\usepackage{latexsym}

\begin{document}

\title{Canonical entanglement for two indistinguishable particles}
\author{XiaoGuang Wang}
\affiliation{Australian Centre of Excellence for Quantum Computer Technology,    
\\
Macquarie University, Sydney, New South Wales 2109, Australia.}
\affiliation{Zhejiang Institute of Modern Physics, Department of
Physics, Zhejiang University, HangZhou 310027, P.R. China }
\author{Barry C. Sanders}
\affiliation{Australian Centre of Excellence for Quantum Computer Technology,    
\\
Macquarie University, Sydney, New South Wales 2109, Australia.}
\affiliation{Institute for Quantum Information Science, University
of Calgary, Alberta T2N 1N4, Canada.}
\date{\today}
\begin{abstract}
We determine the degree of entanglement for two indistinguishable particles
based on the two-qubit  tensor product structure, which is a framework for 
emphasizing
entanglement founded on observational quantities. Our theory connects
canonical entanglement and entanglement based on occupation number
for two fermions and for two bosons and shows that
the degree of entanglement, based on linear entropy, is closely
related to the correlation measure for both the bosonic and fermionic cases.
\end{abstract}
\maketitle

\emph{Introduction.--}
Entanglement and the indistinguishability of particles are two remarkable features 
of
quantum mechanics, yet combining the two in order to obtain a meaningful
quantification of entanglement for indistinguishable particles is challenging.
Several definitions of correlation and entanglement have been
introduced~\cite{Sch01,Pas01,Li01,Zan02,Git02,Wiseman03,Ghi04},
which both illustrates the choices that are available in quantifying entanglement
and underpins the ambiguities about what constitutes the best measure of
entanglement.

If entanglement is to be associated with observational properties that can 
overcome
the indistinguishability of the particles, then the tensor product structure 
(TPS)~\cite{Zan01}
plays a key role. In fact entanglement itself is a consequence of
the TPS and the superposition principle.
However, for systems consisting of many indistinguishable
particles such as bosons and fermions, the TPS is rather subtle.
Approaches to quantifying entanglement can be subdivided into two main
Categories: one based on the canonical
decomposition~\cite{Sch01,Pas01,Li01} and the other based on
the occupation-number representation~\cite{Zan01,Zan02}.
Here we employ both approaches to determine entanglement
for two indistinguishable particles. Our approach is built on two steps:
(i)~to use the canonical
decomposition to obtain the canonical form of the two-particle
states and (ii)~to define entanglement from the canonical
form of the state by the approach based on the occupation-number
representation. We refer to this form of entanglement for indistinguishable
particles as {\em canonical entanglement}. Although large numbers of
particles can be considered, the method is given for two particles, which
can be used as a primitive to treat multiple particles.

\emph{Tensor product structure for states of two indistinguishable particles.--}
A pure state of two particles can be written as
\begin{equation}
|\Psi\rangle=\sum_{i,j=1}^M \Omega_{ij}\hat{a}_i^\dagger \hat{a}_j^\dagger 
|0\rangle,
\label{eq:twoparticles}
\end{equation}
with $\hat{a}_i^\dagger$ and $\hat{a}_j^\dagger$ creation operators for modes~$i$
and~$j$, respectively, and $|0\rangle$ the vacuum state (no particles at all).
For the case of fermions, $\Omega$ is an antisymmetric matrix and the creation
operators are fermionic, while
for the case of bosons, $\Omega$ is a symmetric matrix and the creation operators 
are
bosonic.

For the case of fermions, we will assume that there is an even number of modes
and set $M\equiv 2N$; for the bosonic case the number of modes is not important, 
and we
will assign $M=N$. The canonical form for the two-particle states can be obtained
with the help of the singular value decomposition (SVD)~\cite{Pas01}.
For any antisymmetric $2N \times 2N$ matrix $\Omega^\text{A}\neq 0$,
there exists a unitary operator $U^\text{A}$ such that 
$\Omega^\text{A}=U^\text{A}Y^\text{A}U^{\text{A}T}$, for
$Y^\text{A}\equiv\text{diag}[Y^\text{A}_1,\ldots,Y^\text{A}_N]$ block diagonal 
with blocks
\begin{equation}
Y^\text{A}_i= \left(\begin{array}{cc}
0& y^\text{A}_i \\
-y^\text{A}_i& 0
\end{array}\right).
\end{equation}
and $y^\text{A}_i$ may be zero~\cite{Sch01}. This decomposition is unique and 
yields the
fermionic state
\begin{equation}
\label{eq:PsiF}
|\Psi_F\rangle
    =\sum_{k=1}^N 2y_k \hat{a}_{2k-1}^\dagger \hat{a}_{2k}^\dagger |0\rangle
    =\sum_{k,l=1}^M Y^\text{A}_{kl} \hat{a}_k^{\prime\dagger} 
\hat{a}_l^{\prime\dagger} |0\rangle ,
\end{equation}
for $\hat{a}_k^{\prime\dagger}\equiv\sum_{i=1}^N U_{ik}\hat{a}_i^\dagger$
new fermionic operators. The form above is the
canonical representation~\cite{Everett,Grobe} so we refer to the 
state~(\ref{eq:PsiF})
as being represented in canonical form. The uniqueness of the SVD ensures that the
canonical modes themselves are unique.

Now that we have the canonical form of the two-particle fermion state, we
can impose the TPS~\cite{Zan01} first by establishing the following operators
\begin{align}
\hat{\sigma}_{k+}=&\hat{a}_{2k-1}^{'\dagger} \hat{a}_{2k}^{'\dagger},\,
\hat{\sigma}_{k-}=\hat{a}'_{2k-1}\hat{a}'_{2k},\nonumber\\
\hat{\sigma}_{kz}=&\frac{1}{2}\left(\hat{a}_{2k-1}^{'\dagger}\hat{a}'_{2k-
1}+\hat{a}_{2k}^{'\dagger}\hat{a}'_{2k}-1\right),
\end{align}
which obey su(2) commutation relations.
As operators with different subscripts~$k$ commute with each
other, the state is effectively comprised of distinguishable particles.
Furthermore, the state can be regarded as an $N$-qubit
system with the $k^\text{th}$ qubit state given by the vacuum state
$|0\rangle_k$ and $|1\rangle_k=\hat{a}_{2k-1}^{'\dagger}
\hat{a}_{2k}^{'\dagger}|0\rangle$; the TPS is now evident:
\begin{align}
|\Psi_F\rangle=\sum_{k=1}^N &2y^\text{A}_1|100\ldots 0\rangle
+2y^\text{A}_2|010\cdots 0\rangle+\cdots\nonumber\\
+&2y^\text{A}_N|000\ldots 1\rangle.
\label{qubits}
\end{align}
Further discussions of relevant mappings from fermions to
qubits can be found in Ref.~\cite{Wu}.

For two bosons, $\Omega$ in Eq.~(\ref{eq:twoparticles}) is a
symmetric complex matrix, and $\hat{a}_i^\dagger$ and $\hat{a}_i$
are bosonic creation and annihilation operators. For any symmetric
$N \times N$ matrix $\Omega^\text{S} \neq 0$, there exists a
unitary operator $U^\text{S}$ such that
$\Omega^\text{S}=U^\text{S}Y^\text{S}U^{\text{S}T}$, with
$Y^\text{S}=\text{diag}[y^\text{S}_1,\ldots,y^\text{S}_N]$ and
$y^\text{S}_i$ possibly zero for some values of~$i$. Applying this
unique decomposition to the bosonic state $|\Psi_B\rangle $
obtained from~(\ref{eq:twoparticles}) yields
\begin{equation}
|\Psi_B\rangle
    =\sum_{k=1}^N y^\text{S}_k \hat{a}_{k}^{'\dagger 2}|0\rangle,
    =\sum_{k,l=1}^N Y^\text{S}_{kl}\hat{a}_k^{'\dagger} \hat{a}_i^{'\dagger} 
|0\rangle
\end{equation}
where
\begin{equation}
\hat{a}_k^{'\dagger}=\sum_{i=1}^N U^S_{ik}\hat{a}_i^\dagger
\end{equation}
are new bosonic canonical operators. The TPS for the two-particle bosonic state
is now clear:
\begin{align}
|\Psi_B\rangle=\sum_{k=1}^N &\sqrt{2}y^\text{S}_1|200\ldots 0\rangle
+\sqrt{2}y^\text{S}_2|020\ldots 0\rangle+\ldots\nonumber\\
+&\sqrt{2}y^\text{S}_N|000\ldots 2\rangle.
\label{qubits2}
\end{align}
If we view the two-boson state $|2\rangle$ as a
one-excitation state $|1\rangle$, this state can be
regarded as a multiqubit state, and its entanglement is
well-defined.

\emph{Correlation measures and average entanglement.--} Pa\v{s}kauskas and You 
proposed a
correlation measure to quantify the degree of entanglement~\cite{Pas01}. After 
deriving
the above results, they first obtain the single-particle density matrix and
then obtain the correlation measure determined by the von Neumann entropy
for this reduced state.
For both cases of two fermions and of two bosons, the reduced density matrix is
given by
\begin{equation}
\rho_{\nu\mu}=\frac{\text{Tr}(\hat{\rho}\hat{a}_\mu^\dagger \hat{a}_\nu)}
{\text{Tr}\left(\hat{\rho}\sum_{\mu=1}^M\hat{a}_\mu^\dagger 
\hat{a}_\mu\right)}=2(\Omega^\dagger\Omega)_{\mu\nu},
\label{eq:rhonumu}
\end{equation}
with $M=2N$ and $\Omega=\Omega^\text{A}$ for fermions and
$M=N$ and $\Omega=\Omega^\text{S}$ for bosons.

The von Neumann entropy can be computed from the matrix elements
of Eq.~(\ref{eq:rhonumu}) to obtain
\begin{align}
S   &=-\text{Tr}[\hat{\rho}\log_2(\hat{\rho})]  \nonumber       \\
    &=\left\{ \begin{array}{ll}
        -1-4\sum_{k=1}^N |y^\text{A}_k|^2\log_2(|(y_k^\text{A})^2|),& \text{for 2 
fermions,}\\
        \sum_{k=1}^N (2|y^\text{S}_k|^2)\log_2(2|y_k^\text{S})^2|),&\text{for 2 
bosons.}
        \end{array}\right.
\end{align}
For an uncorrelated state, the entropy $S=1$ for two fermions, but one encounters
the curious situation that the
entropy is not zero for an uncorrelated state~\cite{Wiseman03}.
Our alternative approach, presented below, remedies this problem and
establishes a strong relation between the correlation measure of
Pa\v{s}kauskas and You and canonical entanglement.
In contrast to the fermionic case, $S=0$ does hold for an uncorrelated two-boson 
state;
below we establish the relation between the von Neumann entropy and canonical 
entanglement
for two bosons.

Rather than employ the von Neumann entropy
for the reduced state of a two-particle system, as explained above,
we employ average entanglement to quantify the global entanglement properties;
average entanglement has been employed effectively for studies
of nonlinear inhomogeneous systems~\cite{Lak03,Wan04,Li03,Sco04}.
From Eq.~(\ref{qubits}), the average entanglement between the $k^\text{th}$ 
Fermionic qubit and
the rest.
The average enanglement for two fermions is quantified by the von Neumann entropy
\begin{equation}
	E_{F,k}=h\left(4|y^\text{A}_k|^2\right)
\end{equation}
for
\begin{equation}
	h(x) \equiv -x\log_2 x-(1-x)\log_2(1-x).
\end{equation}
Thus, the average von Neumann entropy is given by
\begin{equation}
	E_F	= \frac{1}{N}\sum_{k=1}^N E_{F,k}
		= \frac{1}{N}\sum_{k=1}^N h\left(4|y^\text{A}_k|^2\right).
\label{EF}
\end{equation}
For a non-entangled state, the entanglement measure $E_F=0$. Moreover,
$E_F$ can be used to quantify the entanglement of two
fermions.  From Eqs.~(\ref{eq:PsiF}) and~(\ref{EF}), we find a relation
between $E_F$ and $S_F$, namely
\begin{equation}
E_F=\frac{1}{N}\left[S_F-1-\sum_{k=1}^N(1-4|y^\text{A}_k|^2)
    \log_2(1-4|y^\text{A}_k|^2)\right].
\end{equation}
The above equation shows that the entanglement definition for two fermions is 
closely
related to the correlation measure.

For two bosons, the entanglement between the $k^\text{th}$ qubit and the remaining
$N-1$ qubits is given by
\begin{equation}
	E_{B,k}=h\left(2|y^\text{S}_k|^2\right).
\end{equation}
and the average entanglement is
\begin{align}
	E_B=&\frac{1}{N}\sum_{k=1}^N
		h\left(-2|y^\text{S}_k|^2\right) \nonumber\\
	=&\frac{1}{N}\left[S_B-\sum_{k=1}^N (1-2|y^\text{S}_k|^2)\log_2(1-
		2|(y^\text{S}_k)^2|)\right].
\end{align}
For a non-entangled state, $E_B=S_B=0$. We see that our measures of
entanglement for two indistinguishable particles are closely related
to the correlation measures for both cases of bosons
and fermions. To reveal a more direct connection between our
entanglement measure and the correlation measure, we next adopt
the linear entropy as a entanglement measure, which is simpler to calculate and 
manipulate
than the von Neumann entropy.

\emph{Linear entropy for the measure of entanglement.--}
Various entanglement measures are used for
different purposes. Above we employ the entropy of
entanglement, which is related to another measure of entanglement,
the squared concurrence (also referred to as the tangle)
$\tau$~\cite{Conc,Jaeger} via the relation
\begin{equation}
E=h\left(\frac{1+\sqrt{1-\tau}}{2}\right).
\end{equation}
For bipartite entanglement of two qubits, the tangle simply
relates to another entanglement measure, linear entropy $E'$,
via $\tau=2E'$. Now we use linear entropy to quantify the
correlation and entanglement. The linear entropy for a state
$\hat{\rho}$ is defined as  $E'\equiv1-\text{Tr}(\hat{\rho}^2)$,
which is simpler to compute than the von Neumann entropy and can also
be a good estimator for the von Neumann entropy~\cite{Ber03}.
As we are considering pure states, the choice of different entropies does
not change the qualitative properties of entanglement.

From Eq.~(\ref{eq:rhonumu}), the two-fermion quantum correlation quantified by the
linear entropy is
\begin{equation}
	S'_F=1-8\sum_{k=1}^N |y^\text{A}_k|^4.
\label{eq1}
\end{equation}
From Eq.~(\ref{qubits}), the entanglement between the $k^\text{th}$ qubit
and the rest is given by
\begin{equation}
	E'_{F,k}=8\left(|y_k|^2-\sum_{\ell=1}^N 4|y_\ell|^4\right).
\end{equation}
Then, the average linear entropy is obtained as
\begin{equation}
	E'_F=\frac{1}N\sum_{k=1}^N \tau_k =\frac{2}{N}\left(1-\sum_{k=1}^N
16|y_k|^4\right). \label{eq2}
\end{equation}
We use the average linear entropy to quantify the entanglement of
two fermions.  From Eqs.~(\ref{eq1}) and~(\ref{eq2}), we obtain
\begin{equation}
	E'_F=\frac{2}{N}(2S'_F-1).
\end{equation}
Thus, the entanglement measure $E'_F$ is proportional to the
correlation measure $S'_F$ up to an additive constant. This result
is important as we can now claim that the correlation of two fermions
considered by Pa\v{s}kauskas and You can be viewed as
entanglement.

For two bosons, the correlation measure
quantified by the linear entropy is
\begin{equation}
	S'^B=1-4\sum_{k=1}^N |y^\text{S}_k|^4.
\end{equation}
From Eq.~(\ref{qubits2}), the average entanglement is given by
\begin{equation}
E'_B=\frac{2}{N}\left( 1-4\sum_{k=1}^N |y^\text{S}_k|^4\right).
\end{equation}
It is evident that the two measures are connected by the following relation
\begin{equation}
E'_B=\frac{2}{N} S'_B ;
\end{equation}
i.e.\ the entanglement measure $E'_B$ is exactly proportional to
the correlation measure $S'_B$. In other words, the entanglement
and the correlation measures are equivalent up to a multiplicative
factor if we adopt the linear entropy to quantify them.

\emph{Conclusions.--} In conclusion, we have given entanglement
measures of two indistinguishable particles, and both cases of
bosons and fermions are considered. The approach here combines the
advantages of the approach based on the canonical decomposition and
another one based on the occupation-number basis. We also exploit the
concept of average entanglement, characterizing the global
entanglement properties of the system.

We compare the entanglement measure with the correlation measure,
and find they are related. Specifically, we find that if we adopt
linear entropy to quantify entanglement and correlation, the
entanglement measure for two fermions is proportional to the
corresponding correlation measure up to a additive constant, and
the entanglement measure for two bosons is equivalent to the
correlation measure up to a multiplicative constant. The
correlation of two fermions considered by Pa\v{s}kauskas and You
can be viewed as entanglement, and this relationship, in turn, supports 
our choices of entanglement measures.

Although we restricted ourselves to the two-particle cases, our
approach shed new lights on quantification of entanglement of
indistinguishable many-body systems. The TPS is the first premise
of quantum entanglement, and thus we have to identify a TPS in
indistinguishable systems in order to define entanglement. The
various TPSs give rise to different measures of entanglement, which can lead
to ambiguities. However, this ambiguity is really an indication of
the complexity of entanglement, associated with the variety of purposes
for which entanglement is useful.

\emph{Acknowledgments.--} This project has been supported by an Australian
Research Council Large Grant. BCS appreciates support from
Alberta's Informatics Circle of Research Excellence (iCORE) 
And the Canadian Institute for Advanced Research (CIAR).

\end{document}